\documentclass[12pt]{article}
\pdfoutput=1
\usepackage{putex}
\usepackage{feyn}
\usepackage[vcentermath]{youngtab}
\usepackage{subfig}
\usepackage{lscape}

\usepackage{graphicx}
\usepackage{epstopdf}
\usepackage{enumerate}
\usepackage{cite}
\usepackage{tensor}
\usepackage{slashed}
\usepackage{amsmath}
\usepackage{amssymb}
\usepackage{mathrsfs}
\usepackage{lgrind}

\usepackage{bbm}

\usepackage{hyperref}

\numberwithin{equation}{section}

\newcommand {\be} {\begin {equation}}
\newcommand {\ee} {\end {equation}}

\newcommand {\bes} {\begin {equation*}}
\newcommand {\ees} {\end {equation*}}

\def\CH{{\cal H}}

\newcommand{\beq}{\begin{equation}}
\newcommand{\eeq}{\end{equation}}

\def\be{ \begin{equation} }
\def\ee{ \end{equation} }

\begin{document}

\preprint{PUPT-2525}

\institution{PU}{Department of Physics, Princeton University, Princeton, NJ 08544}
\institution{PCTS}{Princeton Center for Theoretical Science, Princeton University, Princeton, NJ 08544}

\title{ On Large $N$ Limit of 
Symmetric Traceless Tensor Models
}

\authors{Igor R.~Klebanov\worksat{\PU,\PCTS} and Grigory Tarnopolsky\worksat{\PU}}

\abstract{For some theories where the degrees of freedom are tensors of rank $3$ or higher, there exist solvable
large $N$ limits dominated by the melonic diagrams. Simple examples are provided by models containing one rank $3$ tensor in the tri-fundamental
representation of the $O(N)^3$ symmetry group. 
When the quartic interaction is assumed to have a special
tetrahedral index structure, the coupling constant $g$ must be scaled as $N^{-3/2}$ in the melonic large $N$ limit. In this paper we consider the combinatorics of a large $N$ theory of
one fully symmetric and traceless rank-$3$ tensor with the tetrahedral quartic interaction; this model has a single $O(N)$ symmetry group. 
We explicitly calculate all the vacuum diagrams up to order $g^8$, as well as some diagrams of higher order, and find that in the large $N$ limit where $g^2 N^3$ is held fixed
only the melonic diagrams survive. While some non-melonic diagrams are enhanced in the $O(N)$ symmetric theory compared to the $O(N)^3$ one, we have not found any diagrams
where this enhancement is strong enough to make them comparable with the melonic ones. 
Motivated by these results, we conjecture that the model of a real rank-$3$ symmetric traceless tensor possesses a smooth large $N$ limit where 
$g^2 N^3$ is held fixed and all the contributing diagrams are melonic.
A feature of the symmetric traceless tensor models is that some vacuum diagrams containing odd numbers of vertices are suppressed only by $N^{-1/2}$ relative to the melonic
graphs.
}

\date{}
\maketitle


\section{Introduction and Summary} 

Large $N$ tensor models were introduced in the early 1990s \cite{Ambjorn:1990ge,Sasakura:1990fs,Gross:1991hx}
in an attempt to extend the correspondence of large $N$ matrix models and two-dimensional quantum gravity to dimensions higher than two. 
These early papers contained many new insights, including the importance of the particular quartic interaction vertex for rank-$3$ tensors, where every pair of fields have only one
index in common:
\begin{equation}
V_4= {3 g\over 2} \phi^{abc} \phi^{ade} \phi^{f b e} \phi^{f d c}\ .
\label{tetra}
\end{equation}
Integral over the tensor with a quadratic term and this quartic interaction is not well-defined non-perturbatively because
$V_4$ is not bounded from below.
However, it may be formally expanded in powers of $g$; then it
generates dynamical gluing of tetrahedra and 
was viewed as a step towards understanding 3-dimensional quantum gravity. 

\begin{figure}[h!]
                \centering
                \includegraphics[width=10cm]{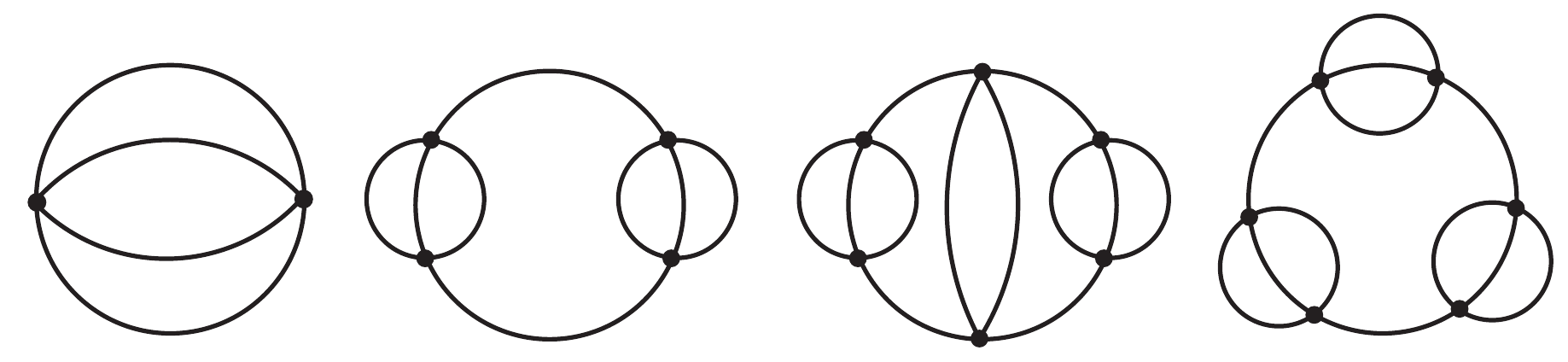}
\caption{All the melonic vacuum diagrams up to order $g^6$.}
                \label{MelonsEx}
\end{figure}

The models considered originally involved tensors with indices transforming under a single symmetry group, $SU(N)$ or $O(N)$,
but the large $N$ limit appeared to be difficult to analyse in such models.  
Years later it was understood that, if the theory has multiple symmetry groups, and the
3-tensors are in tri-fundamental representations, then there is an exactly solvable
large $N$ limit where $g^2 N^3$ is held fixed \cite{Gurau:2009tw,Gurau:2011xp,Gurau:2011aq,Gurau:2011xq,Bonzom:2011zz,Tanasa:2011ur,Bonzom:2012hw,Carrozza:2015adg}. 
Dominant in this limit are the so-called melonic Feynman diagrams (see Figure \ref{MelonsEx}), which are obtained by iterating the insertion of a two-loop sunset
graph into each propagator (this class of diagrams was also studied in the early papers \cite{Patash:1964sp,deCalan:1981chf}). 
The melonic diagrams constitute a small subset of the total number of diagrams (it is considerably smaller than the planar diagrams that dominate in the
't Hooft large $N$ limit \cite{'tHooft:1973jz}, which is used in the matrix models \cite{Brezin:1977sv}), and this accounts for the exact solvability of the theories. Recently there has been a renewed interest in the theories
with tensor degrees of freedom due to their connection \cite{Witten:2016iux,Klebanov:2016xxf} with the SYK-like models of fermions with disordered couplings 
\cite{Sachdev:1992fk,1999PhRvB..59.5341P, 2000PhRvL..85..840G,Kitaev:2015,Polchinski:2016xgd,Maldacena:2016hyu,Jevicki:2016bwu,Gross:2016kjj}.

A class of theories where such a melonic large $N$ limit has been proven to exist have $O(N)^3$ symmetry with a real 3-tensor in the tri-fundamental
representation \cite{Carrozza:2015adg,Klebanov:2016xxf}. In other words, the 3 indices of a tensor are distinguishable, and each one is acted on by a different $O(N)$ group:
\begin{align}
&\varphi^{abc} \to M_{1}^{aa'} M_{2}^{bb'}M_{3}^{cc'}\varphi^{a'b'c'}, \\
& M_{1}\in O(N)_1,\quad M_{2}\in O(N)_2, \quad M_{3} \in O(N)_3\,.
\end{align} 
For such theories one can draw the stranded graphs using the triple-line notation (we may draw each propagator as containing strands of three different colors),
and it is possible to prove the melon dominance.
In particular, all odd orders of perturbation theory are suppressed in the large $N$ limit \cite{Carrozza:2015adg,Klebanov:2016xxf}.
A useful step in the proof is to imagine
erasing all the loops of a given color, i.e. corresponding to one of the $O(N)$ groups, and then counting the remaining loops in the double-line
graphs using their topology. 

Such a method is not available, however, for a theory where there is only one $O(N)$ symmetry group, and the real tensor is in its 3-index irreducible representation
(for example, the fully symmetric traceless one or the antisymmetric one).\footnote{Another interesting model is that of $D$ Hermitian matrices with $U(N)\times O(D)$ symmetry.
Although the standard technique of erasing all the loops of a given color is not applicable to this model, it was argued to be dominated by 
the melonic diagrams in the limit where $N$ and $D$ become large \cite{Ferrari:2017ryl}.}
In \cite{Klebanov:2016xxf} we carried out some perturbative checks of the melonic large $N$ limit in such
tensor models with interaction (\ref{tetra}).\footnote{The counting of $O(N)$ singlet operators in free theories of this type was carried out in \cite{Beccaria:2017aqc}.}
In this paper we report on a complete study of the combinatorial factors of the vacuum diagrams in the theory of a real symmetric traceless tensor
 up to order $g^8$,
as well as some partial results at higher orders. For generating and drawing all diagrams we used the Mathematica program developed in \cite{Kleinert:1999uv}.
We compare with corresponding explicit results for the
theory with $O(N)^3$ symmetry where the real tensor has distinguishable indices. 
We find that the melonic diagrams are dominant in both models. While individual non-melonic diagrams are sometimes enhanced in the $O(N)$ model compared to the $O(N)^3$ model,
these enhancements fall short of making them comparable with the melonic diagrams.

For the vacuum diagrams with even numbers of vertices, the melonic diagrams scale as $ g^{2n} N^{3n+3}$, and we have checked up to $n=4$ that all other diagrams are suppressed
at least by a factor of $1/N$ (we have also checked that this holds for some selected diagrams of order higher than $g^8$). 
These corrections
are present in both the $O(N)^3$ and $O(N)$ models, and there are more contributing diagrams in the latter case.
The vacuum diagrams with odd numbers of vertices behave differently in the two models. The maximum scaling of a graph with
$2n+1$ vertices in the $O(N)^3$ model is 
$ g^{2n+1} N^{3n+3}$, which implies a suppression by $N^{-3/2}$ compared to the melonic graphs. In the $O(N)$ model the maximum scaling is
$ g^{2n+1} N^{3n+4}$, which implies a suppression by only $N^{-1/2}$ compared to the melonic graphs.
Thus, the effective coupling parameter in the $O(N)$ model is of order $N^{-1/2}$, while in the $O(N)^3$ model it is of order $1/N$.
This should have interesting implications for the structure of the large $N$ limit.

Based on our explicit calculations of combinatorial factors, we conjecture that the model of a 3-index symmetric traceless tensor possesses a smooth large $N$ limit where 
$g^2 N^3$ is held fixed and all the contributing diagrams are melonic. As discussed in section \ref{melonicgraphs},
this limit is closely related to the one in the $O(N)^3$ model.

\section{Large $N$ scaling in $O(N)$ and $O(N)^3$ tensor models}

\begin{figure}[h!]
                \centering
                \includegraphics[width=10cm]{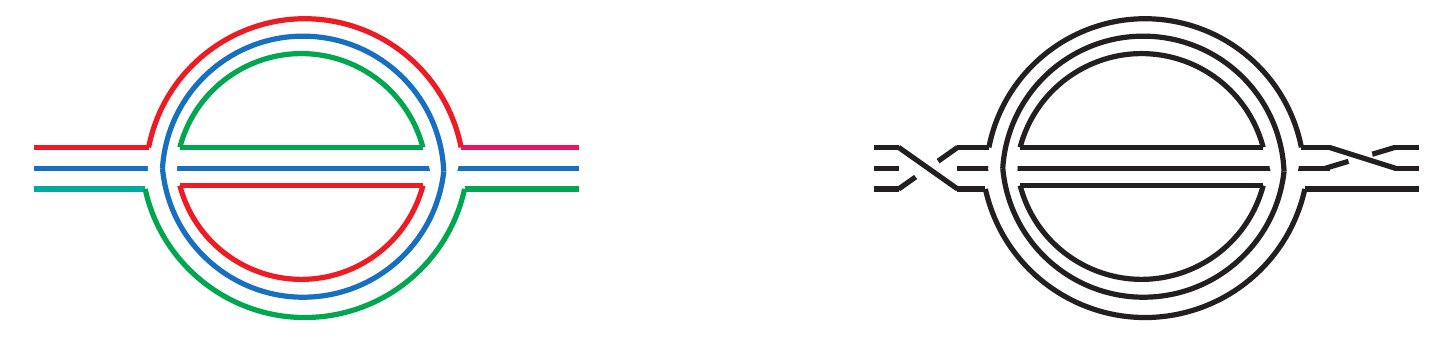}
                \caption{Leading melonic propagator corrections in the $O(N)^3$ and $O(N)$ theories.}
                \label{onemelon}
\end{figure}

In this section we calculate some combinatorial factors for different diagrams in $O(N)^{3}$ and $O(N)$ symmetric theories. 
For the $O(N)^3$ theory we normalize the interaction vertex as
\begin{equation}
{\tilde g\over 4}  \varphi^{a_1 b_1 c_1} \varphi^{a_1 b_2 c_2} \varphi^{a_2 b_1 c_2} \varphi^{a_2 b_2 c_1}\ ,
\end{equation} 
and take the propagator as
\begin{align}
\langle \varphi^{abc}\varphi^{a'b'c'}\rangle_0=\delta^{aa'}\delta^{bb'}\delta^{cc'}\,. \label{O3prop}
\end{align}
The stranded graph for the leading two-loop correction to the propagator is shown in figure \ref{onemelon}. Since there are three index loops (one of each color), this graph
is of order $\tilde g^2 N^3$, and this is the quantity that should be held fixed in the large $N$ limit.\footnote{The correction to two-point function coming from
contracting two fields from the same vertex, i.e. the snail diagram, is of order $g N$. Since this is suppressed in the large $N$ limit where $g\sim N^{-3/2}$, we will
ignore the snail diagrams throughout the paper. Had we not imposed the tracelessness condition on the tensor, there would be diagrams containing multiple
snail insertions which would violate
the melonic limit (we are grateful to F. Ferrari and R. Gurau for pointing this out).}
 More precisely, the two-point function including this graph is
\begin{align}
 \langle \varphi^{abc}\varphi^{a'b'c'}\rangle = \delta^{aa'}\delta^{bb'}\delta^{cc'}(1+    \tilde{g}^2 N^3 +\dots)\,.
\label{melonprop}
\end{align}

In the $O(N)$ model, where the tensor is fully symmetric and traceless, the propagator is
\begin{align}
\langle \phi^{abc}\phi^{a'b'c'}\rangle_0=&\frac {1} {6} \Big (\delta^{aa'}\delta^{bb'}\delta^{cc'}+ \delta^{ab'}\delta^{bc'}\delta^{ca'}+\delta^{ac'}\delta^{ba'}\delta^{cb'}
+ \delta^{ab'}\delta^{ba'}\delta^{cc'}+ \delta^{ac'}\delta^{bb'}\delta^{ca'}+ \delta^{aa'}\delta^{bc'}\delta^{cb'} \notag\\
&-\frac{2}{N+2}\big (\delta^{ab}\delta^{ca'}\delta^{b'c'}+\delta^{ab}\delta^{cb'}\delta^{a'c'}+\delta^{ab}\delta^{cc'}\delta^{a'b'}+
\delta^{ac}\delta^{ba'}\delta^{b'c'}+\delta^{ac}\delta^{bb'}\delta^{a'c'}\notag\\
&\qquad +\delta^{ac}\delta^{bc'}\delta^{a'b'}+\delta^{bc}\delta^{aa'}\delta^{b'c'}+\delta^{bc}\delta^{ab'}\delta^{a'c'}+\delta^{bc}\delta^{ac'}\delta^{a'b'}\big )\Big )
\,. \label{O1prop}
\end{align}
The structure of the melonic two-loop propagator correction in the $O(N)$ model is similar to that in the $O(N)^3$ model (see the stranded
diagrams in figure \ref{onemelon}).\footnote{In figure \ref{onemelon} no two distinct index loops wrap the same cycle of the unstranded $\phi^4$ diagram. This is a general property of the theory with the tetrahedron vertex
 (\ref{tetra}).} 
We again find three additional index loops
which contribute the factor $\sim N^3$. Thus, for the $O(N)$
model a plausible large $N$ limit is with $g^2 N^3$ held fixed. 
More precisely, we find that the leading melonic propagator correction in the $O(N)$ theory is
\begin{align}
 \langle \phi^{abc}\phi^{a'b'c'}\rangle = \langle \phi^{abc}\phi^{a'b'c'}\rangle_0
(1+ g^2 N^3+\dots )\,.
\label{melonpropnew} \end{align}
Note that we have normalized the coupling constant in the $O(N)$ theory as in (\ref{tetra}), which differs by a factor $6$ from the normalization in the $O(N)^3$ theory.
The advantage of this normalization is that the coefficient in (\ref{melonpropnew}) is the same as in (\ref{melonprop}). 

To compute the combinatorial factor of each graph in the $O(N)^3$ theory we represent the tetrahedral vertex as  
\begin{align}
\varphi^{abc} \varphi^{ade} \varphi^{f b e} \varphi^{f d c} = \delta^{aa'}\delta^{bb'}\delta^{cc'}\delta^{dd'}\delta^{ff'}\delta^{ee'}\, \varphi^{abc} \varphi^{a'de} \varphi^{f b' e'} \varphi^{f' d' c'}\,. \label{O3vert}
\end{align}
Then for a given graph, contracting fields using the propagator (\ref{O3prop}) and the $4!$ symmetric configurations of the vertex (\ref{O3vert}) one obtains a sum of products of 
the Kronecker delta symbols. 
Contracting the delta symbols one finds a polynomial in $N$.  
For the $O(N)$ theory the procedure is similar. We may continue to use 
the vertex (\ref{O3vert}) because
the $O(N)$ propagator (\ref{O1prop}) implements symmetrization of the tensor indices.
For example, an explicit evaluation of the melonic vacuum diagram with 2 vertices gives
\begin{align}
\frac{\tilde{g}^2}{8}  (N^6+3N^4+2N^3)
\end{align}
in the $O(N)^3$ model and
\begin{align}
\frac{1}{48} g^2 \frac{(N-2) (N-1) N (N+4) \left(N^5+17 N^4+98 N^3+112 N^2-576 N-768\right)}{(N+2)^3}
\end{align}
in the $O(N)$ model.

The fact that each propagator in the $O(N)^3$ model is made of three strands of different colors make it obvious that two different strands of a propagator
cannot belong to the same loop. In the $O(N)$ model two different strands of a propagator may not belong to the same loop
due to the condition that the tensor is traceless.
Cutting a propagator of a vacuum graph therefore decreases the number of index loops by $3$ and gives a graph contributing to the two-point function.
This shows that each graph contributing to the two-point functions scales as the corresponding vacuum graph times $N^{-3}$.

We would like to prove that the melonic graphs dominate in the large $N$ limit of $O(N)$ theory where $g^2 N^3$ is held fixed.
While we don't know how to do this in general, we have shown that this is the case for
all the vacuum diagrams up to order $g^8$, and 
some selected graphs of higher orders.
We exhibit their pictures and the leading scaling with $N$ in the figures.\footnote{
For each diagram not containing snail insertions, the dominant term at large $N$ is not affected by the 9 terms $\sim 1/(N+2)$ which make the propagator (\ref{O1prop})
traceless. Keeping only the six leading terms in the propagator makes the computer calculation much less time and memory intensive.}
For each diagram the upper integer, shown in black, gives the leading power of $N$ we find in the
$O(N)$ model; the lower integer, shown in blue, gives the leading power we find in the $O(N)^3$ model. If a label B appears below, this means that the diagram is bi-partite, i.e.
it appears in the theory where there are two types of vertices, and each propagator connects different vertices.

In our list of vacuum diagrams we omit the so-called cut vertex diagrams, i.e. the ones that become disconnected if a vertex and the 4 propagators leading to it are erased.
They may also be viewed as (dressed) snail diagrams, i.e. the ones coming from the
``figure eight" graph with the bare propagators replaced by the fully dressed ones.
All such diagrams may be constructed out of a pair of vacuum diagrams by cutting a propagator in each, and then gluing them together using the tetrahedron vertex.
Let us show that this always produces a graph which is suppressed compared to the melonic ones. 
Suppose the two original graphs are of order $g^{m_1} N^{n_1}$ and  $g^{m_2} N^{n_2}$, respectively.
When we cut a propagator in each of the two graphs, we lose a total of 6 index loops (for the symmetric tensor this is true only if the tracelessness is imposed). 
After gluing the two cut graphs into one with the tetrahedron vertex we can recover 4 index loops, but not more. 
For example, in the $O(N)^3$ theory we can make two additional green loops, but only one red and one blue loop
(or an analogous stranded graph with colors permuted). In the theory of a symmetric traceless tensor we can also recover at most 4 index loops.
So, the highest possible scaling of the combined graph is
$N^{n_1+ n_2-2} g^{m_1+ m_2+1}$. 
Even if the two original graphs are melonic, i.e. $n_i= 3+ 3 m_i/2$,
the combined cut vertex graph scales as $N^3 (g N^{3/2})^{m_1+ m_2} (gN)$. It is suppressed by  $N^{-1/2}$ in the melonic limit.\footnote{
If the two cut graphs are glued with the pillow vertex $g_p \varphi^{a_1 b_1 c_1} \varphi^{a_1 b_1 c_2} \varphi^{a_2 b_2 c_2} \varphi^{a_2 b_2 c_1}$,
then we recover 5 index loops. Gluing two melonic graphs in this way gives the cut vertex graph scaling as $N^3 (g N^{3/2})^{m_1+ m_2} (g_p N^2)$.
If $g_p\sim N^{-2}$, then this graph contributes at leading order in the large $N$ limit \cite{Tanasa:2011ur,Carrozza:2015adg}.}

\begin{figure}[h!]
                \centering
                \includegraphics[width=10cm]{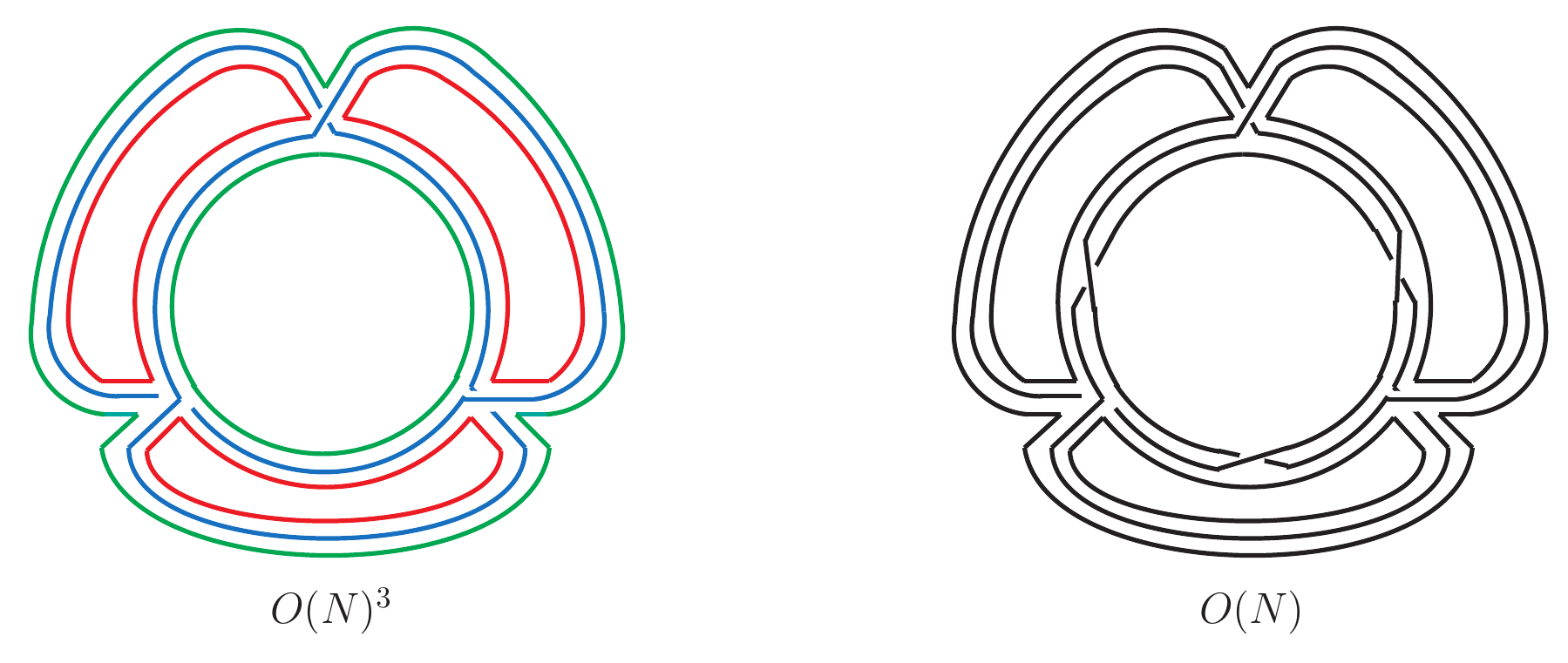}
                \caption{Order $g^{3}$ stranded diagrams for $O(N)^{3}$ and $O(N)$ theories. The diagram in the  $O(N)^3$ theory has 6 loops and scales as $g^3 N^6$;
the diagram in the  $O(N)$ theory has 7 loops and scales as $g^3 N^7$.}
                \label{3ptdiag}
\end{figure}

The first difference in large $N$ scaling between the $O(N)$ and $O(N)^3$ models appears in the diagram of order $g^3$, whose stranded versions are
exhibited in Figure \ref{3ptdiag}.  
The diagram in the  $O(N)^3$ theory has 6 loops and scales as $g^3 N^6$; in the melonic limit this is $\sim N^{-3/2}$.
The diagram in the  $O(N)$ theory has 7 loops and scales as $g^3 N^7$; in the melonic limit this is $\sim N^{-1/2}$. 
These expectations are confirmed by the exact evaluation of the cubic diagram:
\begin{align}
\frac{1}{2} \tilde{g}^3 \left(N^6+3 N^5+3 N^{4}+2N^{3}\right)
\end{align}
in the $O(N)^3$ model, and
\begin{align}
&\frac{1}{24} g^3 \frac{(N-2) (N-1) N (N+4) }{(N+2)^5}\Big(N^8+29 N^7+286 N^6+796 N^5-3120 N^4\notag\\
&\qquad -15232 N^3+12640 N^2+78208 N+58368\Big)
\end{align}
in the $O(N)$ model.
Thus, even though this diagram
is enhanced by $N$ in the $O(N)$ theory, it is still suppressed in the melonic large $N$ limit.

\begin{figure}[h!]
                \centering
                \includegraphics[width=11cm]{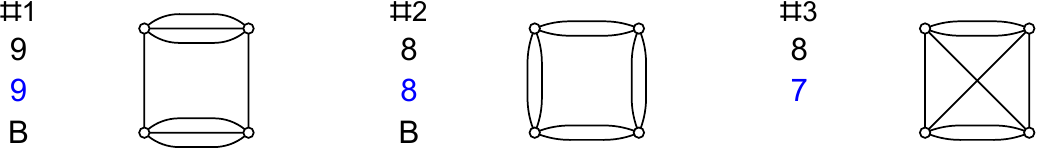}
                \caption{All vacuum diagrams of order $g^4$. The upper integer, shown in black, gives the leading power of $N$ in the
model of a symmetric traceless rank-$3$ tensor of $O(N)$; the lower integer, shown in blue, gives the leading power in the model of a tri-fundamental of $O(N)^3$. 
The letter B labels the bi-partite diagrams.}
                \label{Diags4}
\end{figure}

\begin{figure}[h!]
                \centering
                \includegraphics[width=16.5cm]{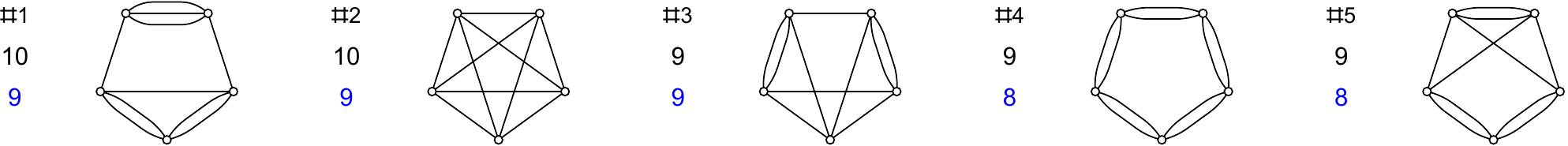}
                \caption{All vacuum diagrams of order $g^5$.}
                \label{Diags5}
\end{figure}

At order $g^5$ there are 5 distinct diagrams, which are shown in Figure \ref{Diags5}. Only the first two are suppressed just by $N^{-1/2}$: diagram $\# 1$ is a melon insertion into the unique graph of order $g^3$, while diagram $\# 2$ (the pentagram inscribed in a circle) is a new strcuture which appears at order $g^5$. 
Interestingly, at order $g^7$ there is no such new structure appearing, so that
the only diagrams suppressed by $N^{-1/2}$ involve melonic insertions into the lower order diagrams.

\begin{figure}[h!]
                \centering
                \includegraphics[width=16.5cm]{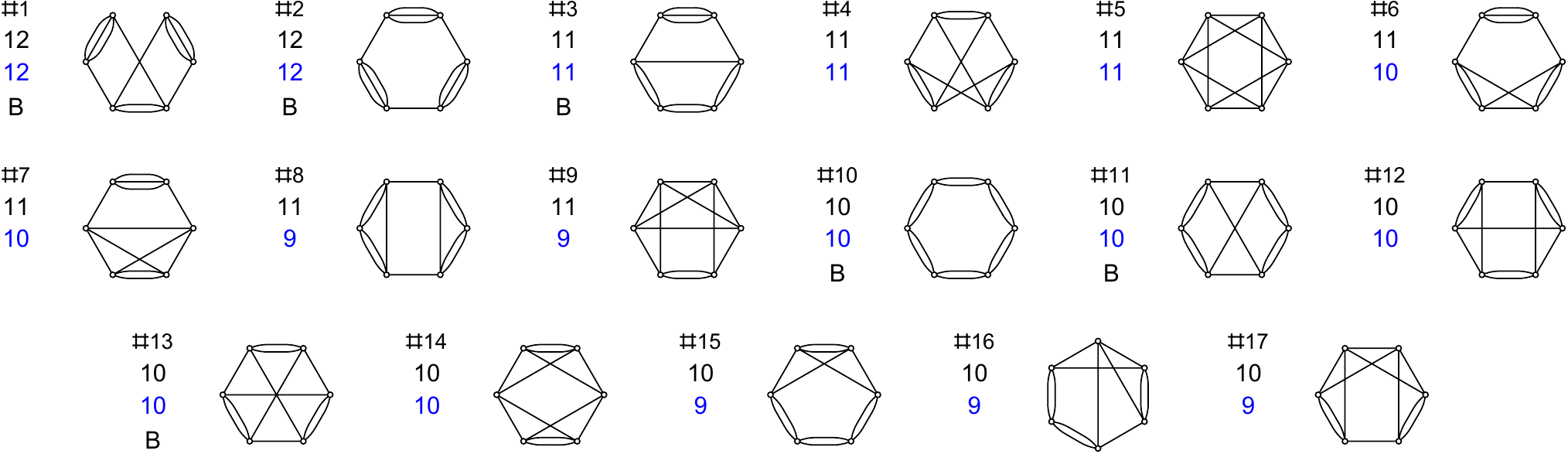}
                \caption{All vacuum diagrams of order $g^6$.}
                \label{Diags6}
\end{figure}

Let us note that some $g^{8}$ graphs in the $O(N)$ theory are enhanced by $N^2$ compared to the $O(N)^3$ case. 
In some cases, this may be traced by to the fact that a diagram with an odd number of vertices, such as the $g^3$ diagram depicted in Fig. \ref{3ptdiag}, may be enhanced by $N$.
Cutting a propagator in such a graph and then gluing two of them gives diagram $\# 8$ of order $g^6$ in Figure \ref{Diags6}; it is indeed enhanced by $N^2$ compared to what is seen in
the $O(N)^3$ theory. However, this diagram is still suppressed by $N$ relative to the melonic ones.
Indeed, in the large $N$ limit where $g\sim N^{-3/2}$ the $g^3$ diagram is suppressed by $N^{-1/2}$ in the $O(N)$ model and by $N^{-3/2}$ in the $O(N)^3$ model.
This translates into suppression of diagram $\# 8$ in Figure \ref{Diags6} by $N^{-1}$ and $N^{-3}$, respectively. 

\subsection{Antisymmetric tensor model}

Another commonly used rank-$3$ representation of $O(N)$ is the fully anti-symmetric one. 
To modify our explicit calculations to the antisymmetric tensor model, we only have to change the index structure of the propagator to
\begin{align}
\langle \phi^{abc}\phi^{a'b'c'}\rangle_0=\frac {1} {6} \big (\delta^{aa'}\delta^{bb'}\delta^{cc'}+ \delta^{ab'}\delta^{bc'}\delta^{ca'}+\delta^{ac'}\delta^{ba'}\delta^{cb'}
- \delta^{ab'}\delta^{ba'}\delta^{cc'}- \delta^{ac'}\delta^{bb'}\delta^{ca'}- \delta^{aa'}\delta^{bc'}\delta^{cb'} \big )\ ,
\end{align}
while the vertex may still be taken to be of the tetrahedral form (\ref{O3vert}). We have carried out extensive perturbative calculations for this model too, and we find that each
individual graph scales with $N$ no faster than in the symmetric traceless model.\footnote{ 
For example, while all graphs of order $g^2$, $g^4$ and $g^6$ have the same leading powers of $N$ as in the symmetric traceless model, 
graph $\# 3$ of order $g^5$ grows $\sim N^8$ in the antisymmetric model compared to $\sim N^9$ in the symmetric traceless model.} 
This provides evidence that the antisymmetric tensor model also has a melonic large $N$ limit.

\begin{figure}[h!]
                \centering
                \includegraphics[width=16.5cm]{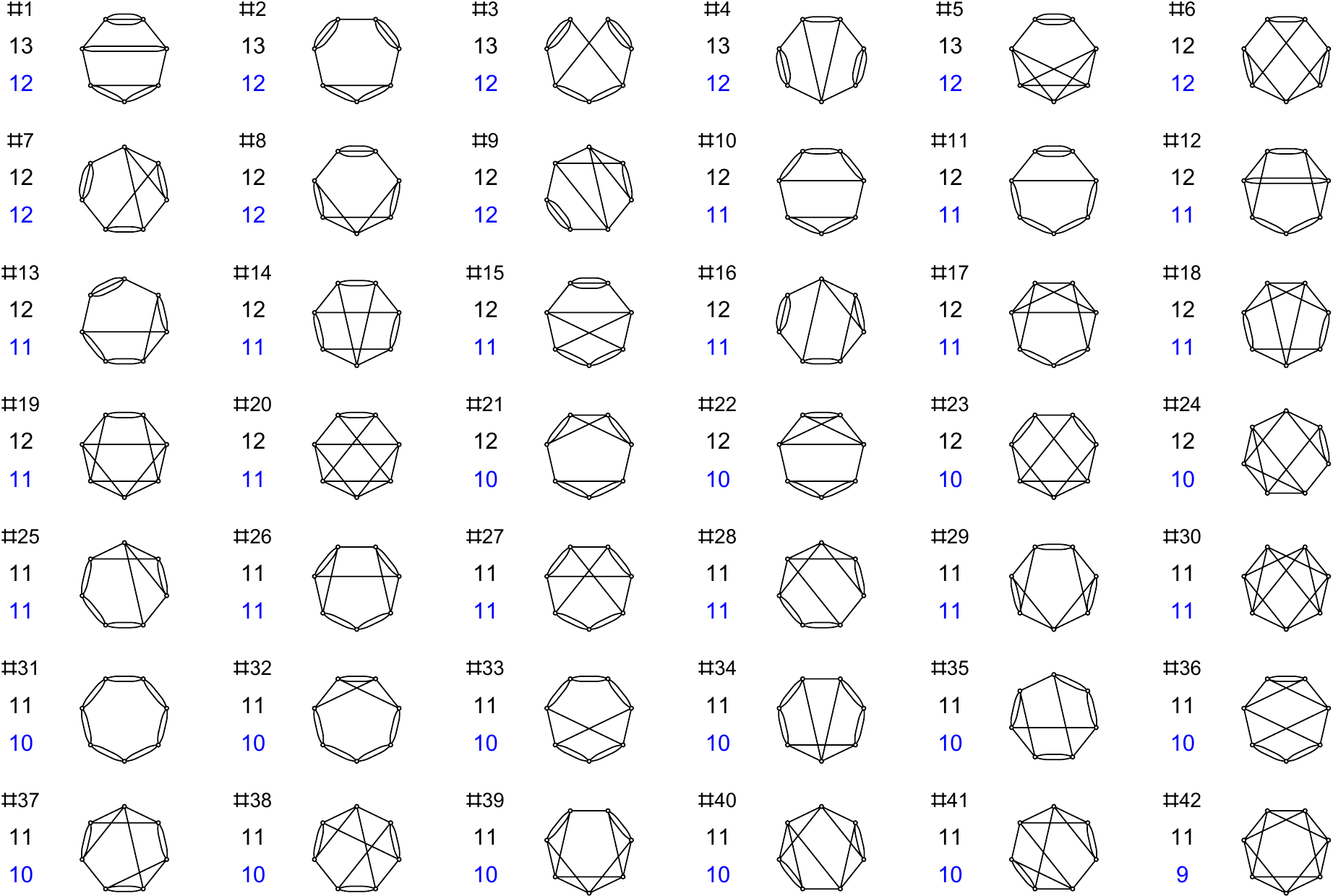}
                \caption{All vacuum diagrams of order $g^7$.}
                \label{Diags7one}
\end{figure}

\begin{figure}[h!]
                \centering
                \includegraphics[width=16.5cm]{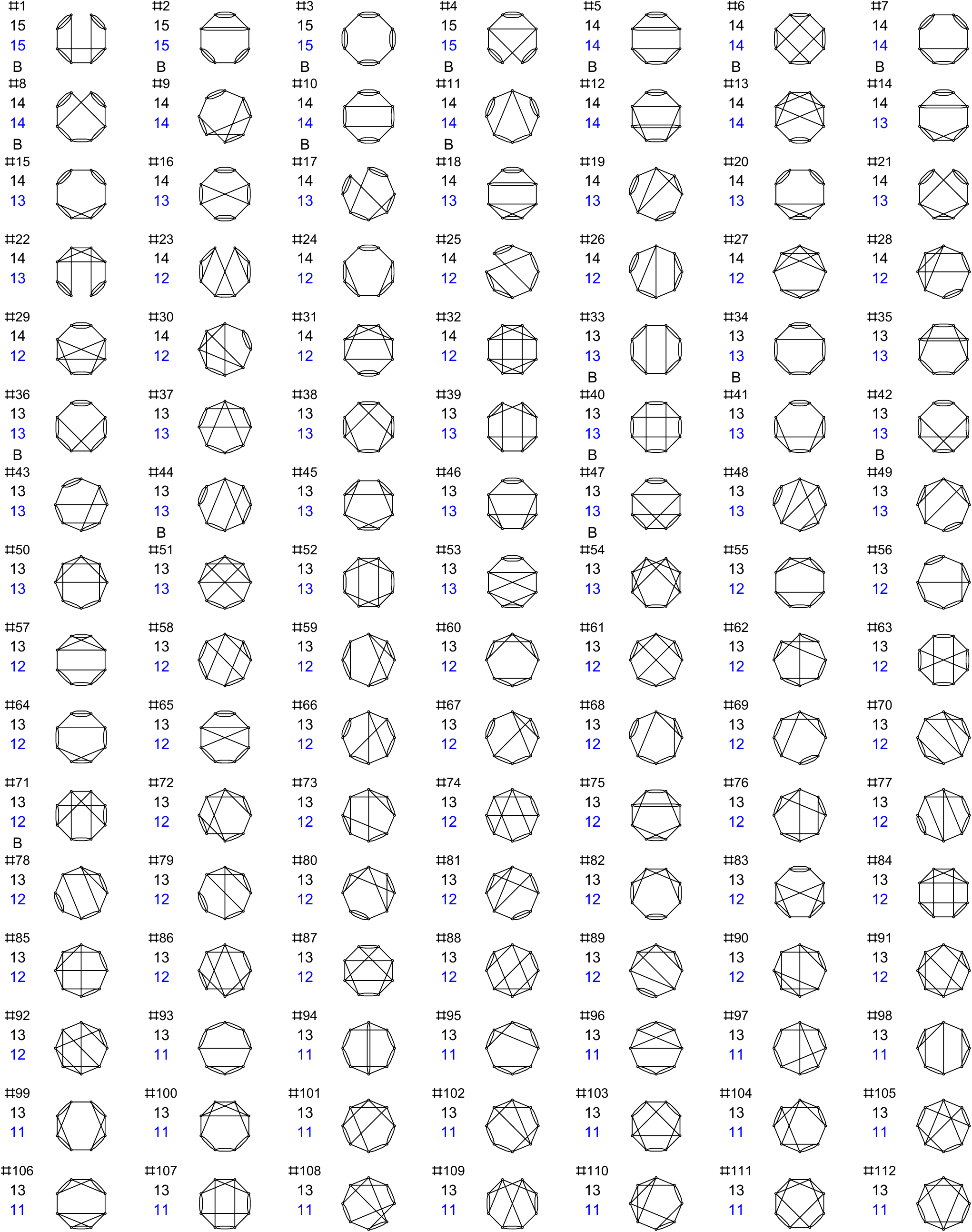}
                \label{Diags8one}
\end{figure} 

\begin{figure}[h!]
                \centering
                \includegraphics[width=16.5cm]{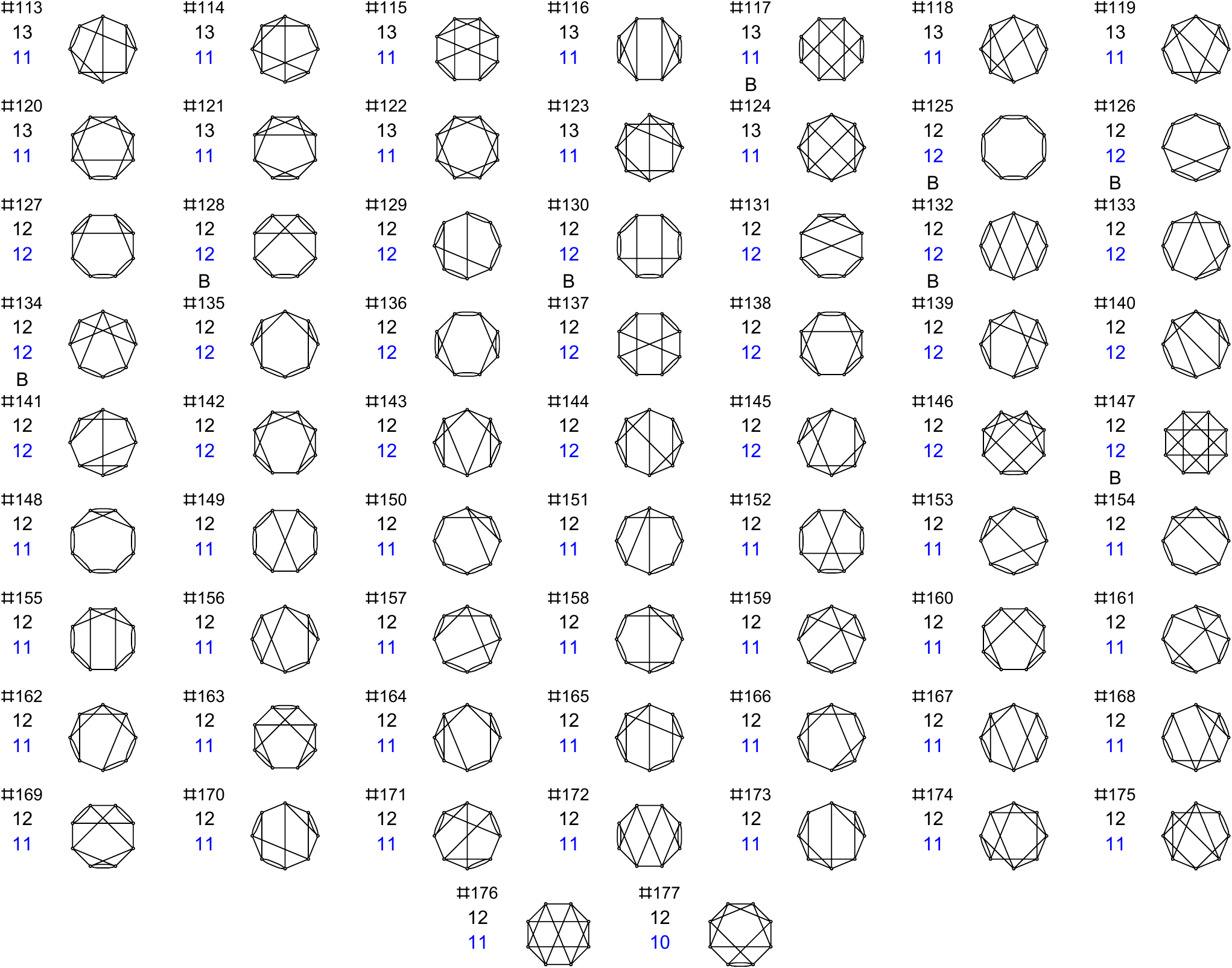}
                \caption{All vacuum diagrams of order $g^8$.}
                \label{Diags8two}
\end{figure}

\section{Bounds on the scaling}

In the models of rank-$3$ tensor each propagator contains 3 strands. The strands are connected into closed loops, and the power of $N$ for each graph is the total number of loops
$n$. If $n_L$ is the number of loops of length $L$, then
\begin{equation}
n= \sum_{L=2} n_L\,,
\end{equation}
where we have excluded loops of length $1$ which can only originate from snail diagrams. Since there are $12$ stranded segments emanating from each quartic vertex, and
each segment connects two vertices, the sum rule on the total number of stranded segments in a graph with $m$ vertices is:
\begin{equation}
\sum_{L=2} L n_L = 6 m\ .
\label{sumrule} 
\end{equation}
The structure of the ``tetrahedral" quartic vertex (\ref{tetra}), 
where every pair of tensors has only one index in common, implies that no closed loop in the Feynman graph can be covered by two
different stranded loops (this would not be the case if the vertex had a pillow structure rather than tetrahedron). This puts an important constraint on the structure of possible
stranded graphs. 
 
For each melonic graph $n= 3 + 3 m/2$. If the theory has a good melonic large $N$ limit, then all other graphs scale with $n< 3+ 3 m/2$.
For the theory with $O(N)^3$ symmetry each strand has a distinct color, and it is possible to perform the counting by erasing one of the colors and relying on the
topology of the double-line graphs. However, such a method is not available for the $O(N)$ theory where the strands are not distinguishable. This implies that there may be
more possibilities for connecting the strands in the $O(N)$ case, so $n_{O(N)}\geq n_{O(N)^3}$. 
The explicit evaluation demonstrates that, for some graphs  $n_{O(N)}- n_{O(N)^3}$ is positive. 
The maximum value of this quantity tends to increase with the order of perturbation theory:
for graphs of order $g^8$ it is 2, while for graph $\# 2$ of order $g^{12}$ it is $4$ (see Figure \ref{Diags12}).

The sum rule (\ref{sumrule}) means that the maximization of $n$ favors graphs with short index loops. As Figure \ref{onemelon} shows, 
each melon insertion into a propagator adds three index loops of length 2, which is hard to beat.\footnote{ 
Our explict results are consistent with the fact that a melon insertion in a graph always increases $n$ by $3$.}
On the other hand, if a Feynman graph contains few faces with perimeter less
than $4$, then  (\ref{sumrule}) leads to a stringent upper bounds on its scaling.
For example, if a graph has no faces with perimeter less than $4$, then
(\ref{sumrule}) implies that $n\leq 3m/2$, which means that the graph is suppressed at least by $N^{-3}$ corresponding to the melonic graphs. 
This inequality is saturated only if $n_L=0$ for $L>4$, i.e. when all index loops have length $4$. 
We notice that the octagram diagram ($\{ 8/3 \}$ in the Schl\" afli notation for polygons), which is number $\# 147$ in Figure \ref{Diags8two}, has no faces with perimeter shorter than 4. 
Our explicit calculation shows that the bound $n\leq 12$ is
saturated for this graph; this means that each stranded loop has lengh $4$.

More generally, if a Feynman graph has $n_2$ distinct faces of perimeter $2$ and $n_3$ distinct faces of perimeter $3$, we find the bound
\begin{equation}
\sum_{L=4} n_L \leq {6 m - 2 n_2- 3 n_3\over 4}\,,
\label{bound} 
\end{equation}
which implies
\begin{equation}
n \leq \frac {n_2} {2}+ \frac {n_3} {4} + \frac{3 m} {2} \, ,
\label{genbound} 
\end{equation}
and the equality may hold only if the RHS is an integer.
A graph may survive in the melonic large $N$ limit 
only if $  \frac {n_2} {2}+ \frac {n_3} {4}$ is $\geq 3$. This is not the case for many non-melonic graphs.

The bound (\ref{genbound}) is often quite informative. For example, for the pentagram graph, which is diagram $\# 2$ in Figure \ref{Diags5}, we find $n_2=0, n_3=10$, so that
$n\leq 10$. The explicit calculation shows that this bound is saturated. As a result, the pentagram graph is suppressed only by $N^{-1/2}$ in the melonic limit.
Moving on to the graphs of order $g^8$,   
for graph $\# 176$ 
we find by inspection that $n_2=0, n_3=4$ so that the bound (\ref{genbound}) is $n\leq 13$. The direct
calculation gives $n=12$, one unit below the bound. For graph $\# 32$  
we find by inspection that $n_2=0, n_3=8$ so that the bound (\ref{genbound}) is $n\leq 14$, and the direct
calculation gives $n=14$. For graph $\# 122$  in Figure \ref{Diags8two} we find by inspection that $n_2=0, n_3=8$ so that the bound (\ref{genbound}) is $n\leq 14$, and the direct
calculation gives $n=13$;  and so on.

The bound  (\ref{genbound}) is particularly easy to apply to the bipartite graphs, which have $n_3=0$. For example,
for graph $\# 6$ of order $g^8$, which is bipartite, $n_2=4$ and the bound is $n\leq 14$. The direct calculation gives $n=14$, so that the bound is saturated. 
Similarly,  for graph $\# 117$ of order $g^8$, $n_2=2$ and the bound is $n\leq 13$; the direct calculation shows that the bound is saturated. 

While the bound (\ref{genbound}) is useful in many cases, it does not provide a proof of the melonic scaling. 
For example, for graph $\# 125$  in Figure \ref{Diags8two}, $n_2=8, n_3=0$, so that the bound (\ref{genbound}) is $n\leq 16$. The actual result $n=12$ is far from saturating
this bound. This is a typical situation for the bubble graphs, of which $\# 125$ is an example. For example, for a bubble graph with $m$ vertices, $n_2=m$, so that the
bound reads $n\leq 4m$. However, the actual scaling is found to be $n=m+4$, which is far less than the bound at large $m$. 

\section{Beyond the eighth order}

A complete study at any order beyond $g^8$ requires calculating the combinatorics of a prohibitive number of graphs, and we have not carried out this task completely.
We have, however, used a combination of direct calculations and the bounds (\ref{genbound}) to make some checks of higher order diagrams.

\begin{figure}[h!]
                \centering
                \includegraphics[width=8cm]{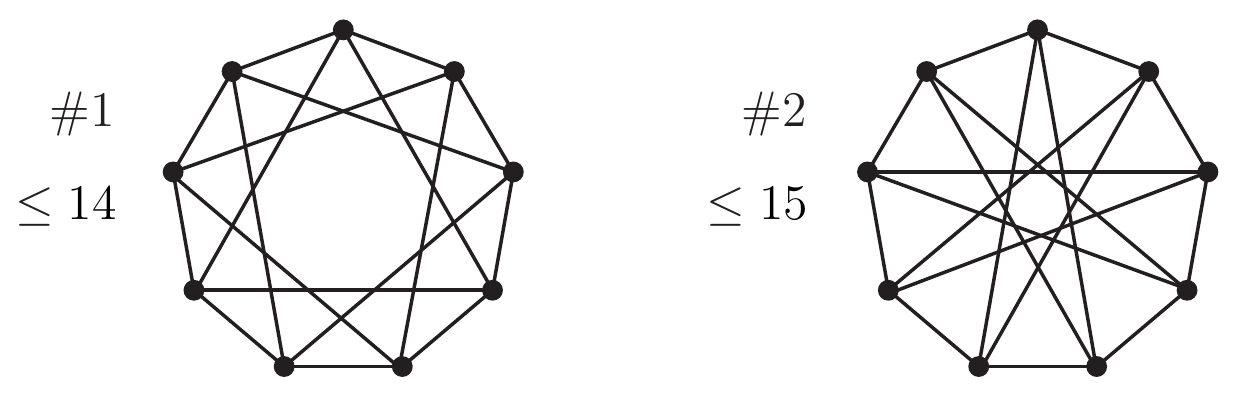}
                \caption{Polygon diagrams $\{9/3\}$ and $\{9/4\}$ inscribed in a circle. The numbers refer to the power of $N$ in the $O(N)$ model.}
                \label{G92111}
\end{figure}

At order $g^9$ one of the most symmetric star shapes is $\{ 9/3 \}$ (in the Schl\" afli notation) inscribed in a circle. This diagram, shown on the left in Figure 
\ref{G92111}, consists of 
three mutually rotated equilateral triangles, and one may wonder if its contriubtion is relatively enhanced similarly to that of the pentagram. However, since $n_2=0$ and
$n_3=3$, we find the bound $n\leq 14$. So, without any direct calculation we see that the diagram is suppressed at least by $N^{-5/2}$ compared to the melonic ones.

For the $\{ 9/4 \}$ inscribed in a circle we have $n_2=0$ and
$n_3=9$, so that the bound (\ref{genbound}) gives $n\leq 15$. This means that
the diagram is suppressed at least by $N^{-3/2}$ compared to the melonic ones.

\begin{figure}[h!]
                \centering
                \includegraphics[width=16cm]{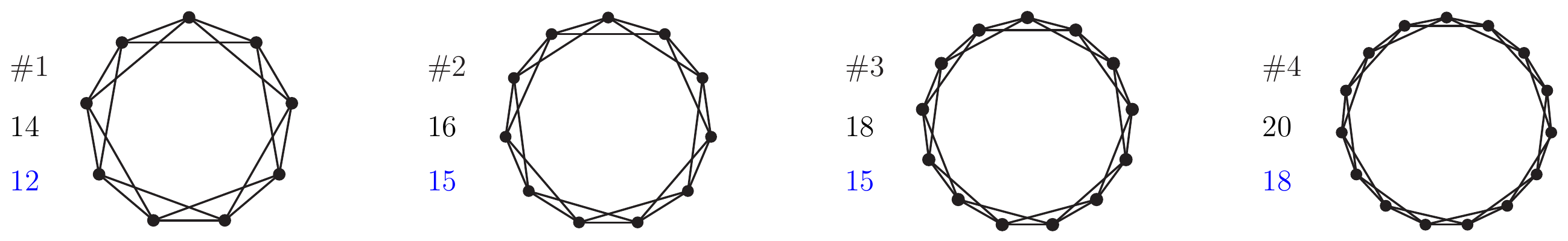}
                \caption{A family of polygon diagrams $\{m/2\}$ inscribed in a circle for $m=9,11,13,15$. The numbers refer to the power of $N$ in the $O(N)$ model.}
                \label{G92112}
\end{figure}

We have also studied the class of polygons $\{ m/2 \}$ (in the Schl\" afli notation) inscribed in a circle. For $m=5$ this is the pentagram, for $m=7$ it is graph $\# 42$ in Figure  \ref{Diags7one}, and for
$m=9,11,13,15$ the graphs are shown in Figure \ref{G92112}. 
With the exception of $m=7$ we find the result $n=m+5$, which shows a linear growth of the scaling with the number
of vertices, similarly to the bubble graphs (for $m=7$ we instead find a smaller value $n=11$). 

\begin{figure}[h!]
                \centering
                \includegraphics[width=16cm]{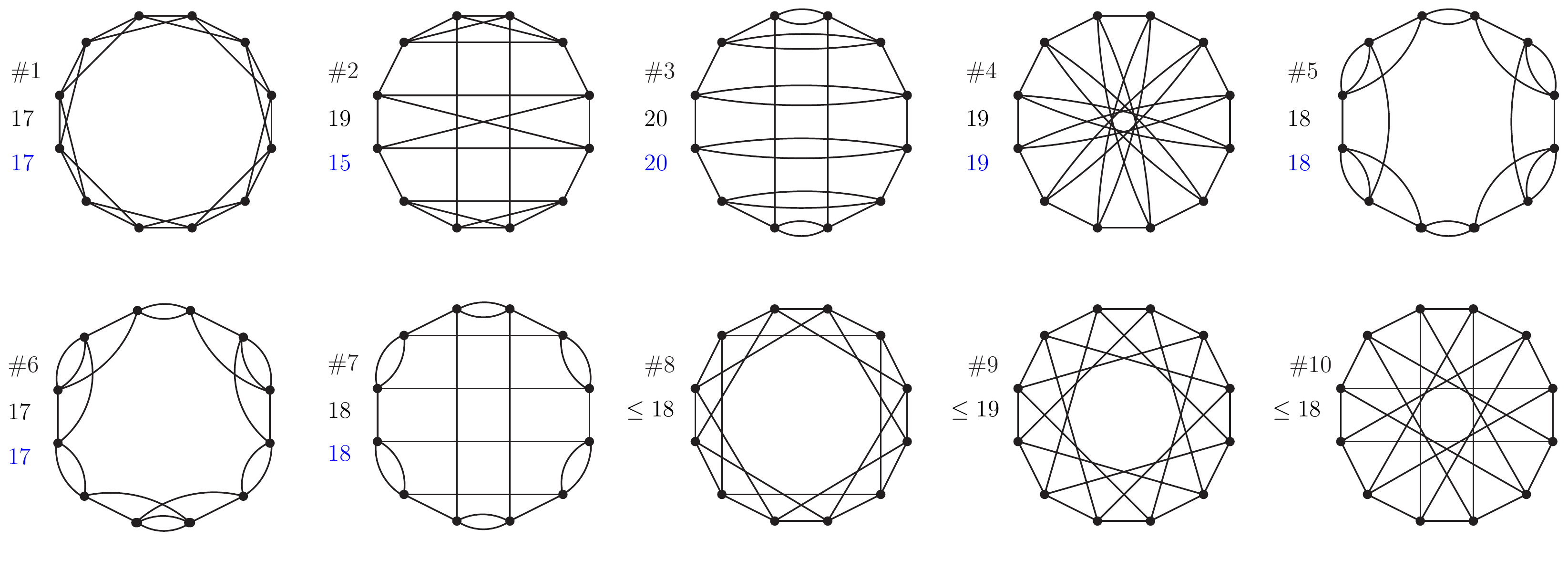}
                \caption{Some non-melonic $g^{12}$ diagrams. They are all suppressed compared to the melonic ones which scale as $N^{21}$.}
                \label{Diags12}
\end{figure}

We have also checked a few graphs at order $g^{12}$; see Figure \ref{Diags12}. 
None of these non-melonic graphs scale as fast as the melonic graphs, which are $\sim g^{12} N^{21}$.
Some graphs, like $\# 8,\# 9,\# 10$, did not need to be calculated explicitly because
the upper bound (\ref{genbound}) shows they are not competitive with the melonic ones. For graph $\# 2$ we observe an enhancement by $N^4$ compared to the
$O(N)^3$ theory, but the graph is still suppressed by $N^{-2}$ compared to the melonic ones.

\section{Melonic graphs} 
\label{melonicgraphs}

Let us define normalized interaction terms in the $O(N)^{3}$ and $O(N)$ cases 
\begin{equation}
V_{O(N)^{3}} = \frac{1}{4}\tilde g \varphi^{abc} \varphi^{ade} \varphi^{f b e} \varphi^{f d c} , \qquad V_{O(N)} =\frac{3}{2}g \phi^{abc} \phi^{ade} \phi^{f b e} \phi^{f d c}\,,
\end{equation}
where the rank 3 tensor field $\varphi^{abc}$ has distinguishable indices, while $\phi^{abc}$ is a symmetric traceless tensor.
For the $O(N)^3$ theory the sum over connected melonic vacuum graphs  in the large $N$ limits is
 \begin{equation}
F_{O(N)^3}= N^3 (\sum_{n=1}^\infty a_{2n} \tilde \lambda^{2n} )\ ,
\label{vacgraphs3} 
\end{equation}
where $\tilde \lambda^2= \tilde g^2 N^3$. The specific coefficients $a_{2n}$ depend on the dimensionality and the field
content of the theory.
For example, for a scalar theory in $d=0$,
\begin{equation}
a_2=\frac {1} {8} \ , \quad a_4= \frac {1} {4} \ , \quad a_6= \frac {11} {12} \ , \quad a_8= \frac {35} {8},\quad \ldots ,\quad a_{2n}= \frac{1}{8n(4n+1)}{4n+1 \choose n}\,.
\end{equation}
These coefficients can be obtained by solving Schwinger-Dyson equation for the two-point function in the $d=0$ dimension  \cite{Bonzom:2011zz}
\begin{equation}
G_{\textrm{melons}}(\lambda) = 1+ \lambda^{2}G_{\textrm{melons}}(\lambda)^{4}\,.
\end{equation}
Then free energy $F$ is obtained from $G_{\textrm{melons}}$ through the relation $G_{\textrm{melons}} = 1+4\lambda \partial_{\lambda}F/N^3$.

Now, if we assume that in the large $N$ limit melonic graphs dominate also in the $O(N)$ model, then we 
expect to find the same expression in terms of the coupling $ \lambda^2=  g^2 N^3$, up to an overall factor:
\begin{equation}
F_{O(N)}= \frac {N^3} {6}  (\sum_{n=1}^\infty a_{2n}  \lambda^{2n} )\,.
\label{vacgraphs3} 
\end{equation}
The reason for the factor $1/6$ is that the number of degrees of freedom in the symmetric traceless rank $3$ tensor is $N(N+4)(N-1)/6= N^3/6 + O(N^2)$.
We explicitly checked (\ref{vacgraphs3}) up to order $\lambda^8$. 
 So, the melonic limits in the $O(N)$ and $O(N)^3$ models are simply related.

\section*{Acknowledgments}

We thank F. Ferrari and R. Gurau for very useful discussions and comments on a draft of this paper, which pointed us towards the necessity of imposing the 
tracelessness condition on the symmetric tensor.
We are also grateful to E. Witten for very useful discussions and encouragement, and to F. Schaposnik Massolo for useful correspondence.
The symbolic calculations were carried out using Mathematica on the Princeton University Feynman computer cluster.  
The work of IRK and GT was supported in part by the US NSF under Grant No.~PHY-1620059. GT also acknowledges the support of a Myhrvold-Havranek Innovative Thinking Fellowship.


\bibliographystyle{ssg}
\bibliography{Diagrams}

\end{document}